\documentclass[10pt,twocolumn,english,aps,reprint]{revtex4}
\usepackage[T1]{fontenc}
\usepackage[latin9]{inputenc}
\usepackage{color}
\usepackage{amstext}
\usepackage{amssymb}
\usepackage{graphicx}

\makeatletter
\@ifundefined{textcolor}{}
{%
 \definecolor{BLACK}{gray}{0}
 \definecolor{WHITE}{gray}{1}
 \definecolor{RED}{rgb}{1,0,0}
 \definecolor{GREEN}{rgb}{0,1,0}
 \definecolor{BLUE}{rgb}{0,0,1}
 \definecolor{CYAN}{cmyk}{1,0,0,0}
 \definecolor{MAGENTA}{cmyk}{0,1,0,0}
 \definecolor{YELLOW}{cmyk}{0,0,1,0}
}

\makeatother

\usepackage{babel}
\begin{document}

\title{$\delta$-doped LaAlO$_{3}$-SrTiO$_{3}$ interface: Electrical transport and characterization of the interface potential}

\author{A. Rastogi$^{1}$, S. Tiwari$^{2}$, J. J. Pulikkotil$^{3}$, Z. Hossain$^{1}$, D. Kumar$^{2}$
and R. C. Budhani$^{1,3}$}

\affiliation{$^{1}$Condensed matter- Low Dimensional Systems Laboratory, Department
of Physics, Indian Institute of Technology, Kanpur 208016, India\\
$^{2}$School of Physical Sciences, Jawaharlal Nehru University, New
Delhi 110067, India\\
$^{3}$CSIR- National Physical Laboratory, Dr. K. S. Krishnan Marg,
New Delhi 110012, India}

\begin{abstract}
Here we investigate LaAlO$_{3}$-SrTiO$_{3}$ heterostructure with $\delta$-doping of the interface by LaMnO$_{3}$ at less than one monolayer. This doping strongly inhibits the formation of mobile electron layer at the interface. This results in giant increase of the resistance and the thermopower of the heterostructure. Several aspects of this phenomena are investigated. A model to calculate the carrier concentration is presented and effect of doping and detailed temperature dependence is analyzed in terms of model parameters and the weak-scattering theory. The large enhancement of thermopower is attributed to the increased spin and orbital entropy originating from the LaMnO$_{3}$ mono-layer.
\end{abstract}
\maketitle

The recent discovery of a highly mobile two-dimensional electron liquid (2-DEL)
at the interface of two perovskite oxides has naturally evoked a huge amount of
interest\cite{Ohtomo04,Thiel06,Nakagawa06}. These heterostructures have
since been widely investigated both experimentally
and theoretically, (a) to understand the mechanism of formation of 2-DEL
\cite{Herranz07,Willmott07,Lee08,Reinle12} (b) to investigate its properties and examine
new phenomena exhibited by it \cite{Reyren07,Basletic08,Biscaras10,Janicka} and (c)
for exploitation of its unique properties for technological ends \cite{Hwang12,Mannhart11}.
Here we contribute to this effort by investigating a $\delta$-doped interface. We
have prepared an interface of SrTiO$_3$ and LaAlO$_3$ with $\delta$-doping of
a sublayer of LaMnO$_3$ at the interface. Fig. 1a shows the layer structure of
our system.

Our motivation was to seek a material structure with improved thermoelectric
performance. The quantum size effect in low-dimensional systems is now well known
to enhance thermopower \cite{Hicks01,Martin}. Ohtomo et al \cite{Ohtomo02}.
investigated superlattices of SrTiO$_3$ and LaTiO$_3$ and showed
that at interfaces the Ti ion exist in mixed valence states (Ti$^{3+}$ and
Ti$^{4+}$). Such mixed-valent states along with strong correlations have
interesting consequences for the enhancement of TEP, as earlier work
\cite{Chaikin,Marsh,Koshibae01} has shown that large contribution can come
from spin and orbital degrees of freedom.

In the present work, we report electrical transport, the resistance and the thermoelectric
power in a series of LaAlO$_{3}$-SrTiO$_{3}$ oxide hetero-structures with $\delta$-doping
by a fraction of a single monolayer of Mn ions in the vicinity of the interface. Our key finding is that
this sublayer has a drastic effect on the properties of the interface. The resistance and
thermopower increase considerably compared to the undoped interface. These effects are
undoubtedly related to the mechanism of formation of the nearly two-dimensional electron liquid (2-DEL)
at the interface. We find that our set of measurements offer an opportunity to develop a simple
model which extends the standard polar-catastrophe ideas. While interface physics is rather complex
due to several factors, like relaxation of ionic positions, resulting electrostatic forces, disorder
and other interface reconstruction effects, our simple analysis of transport properties does pave
way for a better understanding of some key issues of the interface physics.

Pulsed laser deposition (with KrF laser of wavelength $248$ nm) technique
has been employed for layer by layer growth of LaAlO$_{3}$ films
on TiO$_{2}$ terminated (001) oriented SrTiO$_{3}$ single crystal
substrate. To tailor atomically sharp interfaces, growth parameters
such as O$_{2}$ partial pressure and substrate temperature have been
critically examined since these parameters determine the film stoichiometry
and defects \cite{Junquera,Takizawa,Yamada}. The
substrate was kept 7 cm away from the target. The laser pulses were
fired at 2 Hz and flunce $\sim$1 J cm$^{-2}$ per pulse, which leads
to a growth rate of $\sim$ 0.15 $\textrm{\AA}$/s. Further details
of growth are described in our earlier publications \cite{Rastogi,Rastogi2}.
A fraction $\delta$ ($0\leq\delta\leq0.6$) of a monolayer of LaMnO$_{3}$
was first deposited on the TiO$_{2}$ terminated SrTiO$_{3}$ substrate
at 10$^{-4}$ mbar of O$_{2}$ pressure and 800 $^{o}$C, followed
by 20 u.c. thick LaAlO$_{3}$ film. Sample was thereafter cooled
under the same deposition pressure. These heterostructures were characterized
by X-ray diffraction on PANalytical X'PERT PRO, revealing a tetragonal
strained perovskite structure. Top electrodes of Ag/Cr were deposited
in Van der Pauw and standard four probe geometry on films using shadow
masking. The thermoelectric measurements were
carried out in the Quantum Design PPMS.

\begin{figure}[h]
\begin{centering}
\includegraphics[width=8cm]{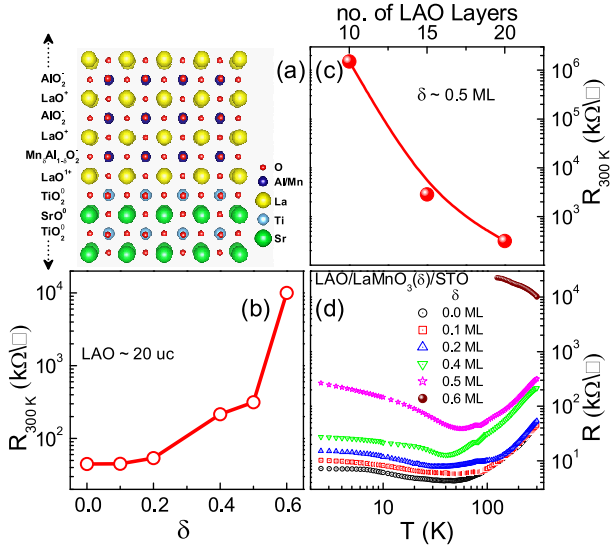}
\par\end{centering}
\caption{\label{RTcurve} Color online: (a) shows interface of STO and LAO with
$\delta$-doping of a fraction of monolayer of LaMnO$_3$. Note that the second layer from the interface
on LAO side is Mn$_{\delta}$Al$_{1-\delta}$. (b) Variation of room temperature R$_{\square}$
of the heterostructure with of $\delta=0.5$ with the thickness of LaAlO$_{3}$
from 10 to 20 u.c.  (c) The change in R$_{\square}$ with increasing
doping at room temperature. (d) Variation of the sheet resistance (R$_{\square}$)
 with temperature of samples with doping fractions $\delta$ from 0.0 to 0.6 in the temperature range
from 5K to 200K. No metallic behavior for $\delta > 0.5$.}
\end{figure}

In order to describe various effects of the Mn-layer on the formation of the interfacial
2-DEL, we begin with the observation that no interfacial electronic conductivity was
observed when LaMnO$_{3}$ films of several unit cells were deposited
on TiO$_{2}$ terminated SrTiO$_{3}$ substrates \cite{Kim}. The interfacial
conduction could be observed only with less than one layer of LaMnO$_{3}$.
This implies a unit cell consisting of one layers Mn$_{\delta}$Al$_{1-\delta}O_2$ and
LaO at the interface as shown in Fig. 1(a). Whereas for LAO-STO system a minimum of
4 to 5 unit cells of LAO are required on TiO$_2$-terminated substrate of STO to form a 2-DEL,
for the doped interface the minimum number of LAO unit cells required to achieve the same
is a lot larger and strongly depends on the Mn concentration in the layer.
Fig.\ref{RTcurve}(b) shows the change of sheet resistance on decreasing
the LAO thickness for $\delta$ = $0.5$.  One sees that R$_{\square}$ at room temperature increases
sharply by three orders in magnitude as the LAO thickness decreases from $20$ to
$10$ u.c. The presence of even half a monolayer of LaMnO$_{3}$
changes the number of LAO layers required to achieve a good conducting 2-DEL, to 17 to 20
as compared to 4 to 5 for STO-LAO interface.
To further confirm this finding, we reduced the LaMnO$_{3}$ to monolayer fraction $\delta$=$0.2$.
Here we find that $\approx$ $13$ - $15$ u.c., were required
to make the interface conducting. Moreover, for $\delta$ $>$ $0.5$,
we find that the system remains insulating for $20$ u.c., of LaAlO$_{3}$.
The variation in R$_{\square}$(T) as a function of LaMnO$_{3}$ monolayer
fraction ($\delta$) for a fixed LaAlO$_{3}$ over-layer ($20$ u.c.)
is shown in Fig.\ref{RTcurve}(c). We see a very rapid rise in resistance
when $\delta$ exceeds $0.2$, confirming again that the LaMnO$_{3}$ monolayer inhibits
the formation of 2-DEL at the interface in a strong way. The Fig.\ref{RTcurve}(d) shows the
variation of the sheet resistance (R$_{\square}$) with temperature in the range 2K to 300K
for five films with monolayer Mn-doping in the range $0\,\leq\delta\,\leq0.5$ and an optimal
LAO thickness of 20 u.c.. The data is also shown for $\delta=0.6$, at which doping the
interface is non-metallic. All the curves show a minimum in
temperature range of 40K to 50K. The decrease from room temperature
to the minimum is rather sharp, the resistance decreasing by a factor of about 7 to 8.
At lower temperatures there is a slow increase of resistance suggesting a weak-scattering regime.
In order to understand this set of results we have developed a simple model which primarily calculates
the density of the electron liquid at the interface as a function of temperature. Referring to Fig. 1(a),
recall that LAO due to its alternating charged layers (LaO)$^+$ and (AlO$_2)^-$ creates a large positive
potential at the interface compared to the ground or surface potential. In the ideal picture layers can
be treated as capacitors in series, the potential offset is
proportional to the number k of such pairs. Recall that for the bulk STO a gap of 1.6 eV exists
between the conduction band and the Fermi level and it is this gap that is presumably reduced
by the LAO layers. Therefore a certain minimum number of unit cells of LAO are needed to provide a
threshold potential to draw electrons to the conduction band of STO. The potential catastrophe with
increasing k is clearly avoided by the solid by relaxation of ionic positions, buckling of layers etc.,
which screen the field. This makes the resulting potential far smaller and saturate rapidly with k.

Another important point to note is that, if the layers are charge-balanced they only produce this potential
offset i.e. potential is constant outside the set of charged layers. To generate
the electron layer at the interface. one needs an electric field which can come from some unbalanced
positive charge, which one can reasonably presume, exists on the LaO layer at the interface.
Electrons drawn to the conduction band of STO occupy mostly the Ti $d_{xy}$ levels
of TiO$_2$ layer at the interface \cite{Delugas}. These electrons
screen the positively charged LaO layer at the interface.
Though there is smaller occupation in other bands like $d_{zx}$ and $d_{yz}$  till around five layers,
the mobile electrons should mostly be from Ti $d_{xy}$ levels
due to their high dispersion along the interface \cite{Popovic,Pentcheva,Delugas}.

This leads us to account for the role of LAO layers by two parameters: a potential offset
$\phi_0$ and an excess charge density $\alpha e/a^2$ (a=lattice parameter of $LaO$ layer).
Due to potential offset, the excitation gap in the $TiO_2$ is reduced to $\Delta=(1.6-e\phi_0)$ eV. The
charge density produces an electric field $E_0 = \alpha e/(2 \epsilon_0 \epsilon_s a^2)$, where
$\epsilon_s$ is the dielectric constant of STO which is also temperature dependent. We now use
Thomas-Fermi theory to find a self-consistent equation for potential on STO side. Thomas-Fermi
theory is valid only when length scale of potential variation is much larger than the De Broglie wavelength,
which approximation is not quite valid near the interface. We assume that
the charge is induced by occupation in the conduction band of STO.  Taking z-axis to be perpendicular
to the interface, the carrier density $N(z)$ is given in terms of potential $V(z)$ as $N(z)= 2g(u(z) +\beta \Delta)/{\lambda_T}^3$. Here $u(z) = \beta e V(z)$, $\beta = (k_BT)^{-1}$ and $\lambda_T = \hbar/(2m^* k_BT)^{1/2}$. where $m^*$ denotes the effective mass in the conduction band of STO.
The function $g(u)$ is essentially the Fermi-Dirac distribution given by
\begin{equation}
 g(u) = \frac{2}{\sqrt \pi}\int_0^{\infty} \frac{x^{1/2} dx}{e^{x+u} + 1}. \nonumber
\end{equation}
This enables us to obtain the equation of potential as
\begin{equation}
\frac{d^2u}{dz^2} = \frac{1}{2 \lambda^2}\left[\frac{g(u+\beta \Delta)}{g(\beta \Delta)}- 1 \right],
\end{equation}
where the screening length $\lambda$ is given by
\begin{equation}
\lambda^{-2} = \frac{4 \beta e^2 g(\beta \Delta)}{\epsilon_0 \epsilon_s \lambda_T ^3},
\end{equation}
The equation is solved for the boundary condition given by the field $E_0$ and the profile of
carrier density is found. The sheet density $N_2(T)$ is obtained by integrating $N(z)$ over a
depth of order $\lambda$. The variation of $N_2$ with temperature for a typical set of parameters
used here is shown in the inset of Fig. 2. It should be mentioned that the sheet density is
a rather sensitive function of parameters $\Delta$ and $\alpha$. These parameters mimic changes
in the electronic density due to the presence of interface and subsequent ionic relaxations. Here they
are obtained from experimental parameters as discussed below.

We first discuss the effect of doping in qualitative terms. As shown in Fig. 1(a),
when the LaMnO$_{3}$ layer is inserted, the Mn ions occupy the negatively charged layer next
to LaO but away from the interface. We surmise that due to mixed-valent character of Mn ion between states
Mn$^{3+}$ and Mn$^{4+}$ a charge transfer occurs from negatively charged Mn$_{\delta}$Al$_{1-\delta}$O$_2$
layer toward the LaO layer neutralizing its charge considerably. Further a reduced charge on Mn- layer
would also lead to a relaxation of charge on other LAO layers, thereby affecting both $\phi_0$
and $\alpha$. Small changes in these parameters can drastically affect
the field on the TiO$_2$ layers at the interface and consequently a larger number of
LaO layers are needed to generate the interfacial 2-DEL. Clearly, an ion with mixed-valent character
greatly facilitates the charge transfer, which was restricted with Al ions.

Next we discuss the detailed temperature dependence of resistance as shown in
Fig. \ref{RTcurve}(d). This allows us to obtain the model parameters which corroborate to an
extent the qualitative reasoning described above. We focus on the weak-scattering regime for
temperatures below the resistance minimum. The resistance in this regime is fitted according
to the formula,
\begin{eqnarray}
%\begin{equation}
\label{resistance}
 \sigma(T) =\frac{N_2(T) e^2 \tau_0}{m^*} - \frac{e^2}{4\pi^2 \hbar}[2 \ln  ( \frac {\tau_0^{-1}
+ \tau_i^{-1} + \tau_{so}^{-1}}{\tau_i^{-1} + \tau_{so}^{-1}}) \nonumber\\
 + \ln ( \frac {\tau_0^{-1}+ \tau_i^{-1} + 2\tau_{so}^{-1}}{\tau_i^{-1} + 2 \tau_{so}^{-1}} )- \ln  ( \frac {\tau_0^{-1} + \tau_i^{-1}}{\tau_i^{-1}})]
%\end{equation}
\end{eqnarray}
The first term is the Drude term and the second one is the weak-scattering correction describing
quantum interferece effects \cite{Hikami,Bergmann,Punnoose}. Three relaxation
times occur in this expression. First is the elastic scattering time $\tau_0$ which is temperature
independent. The second one is the inelastic scattering time $\tau_i$ whose temperature dependence is
taken to be of the form $\tau_0 (T_0/T)^p$ where p depends on the inelastic scattering mechanism. The
inelastic scattering typically gives rise to logarithmic increase of resistance with decreasing
temperature. Finally $\tau_{so}$ is the relaxation time due to spin-orbit scattering. This is important
here due to interface electric field which causes significant effect on electrons moving parallel
to the interface \cite{Caviglia10}. The spin-orbit scattering works oppositely to the inelastic scattering
and it slows down the increase of resistance at lower temperatures, where $\tau_{in} > \tau_{so}$.

\begin{figure}[h]
\begin{centering}
\includegraphics[width=8cm]{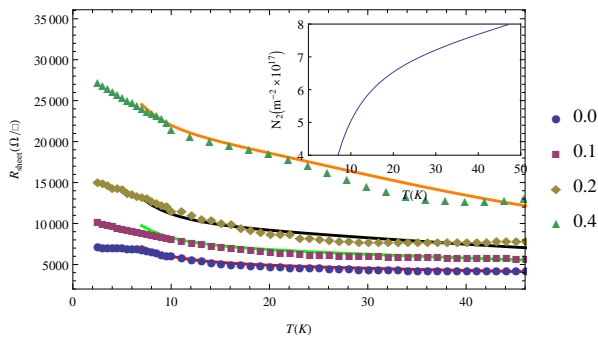}
\par
\end{centering}

\caption{\label{Fits} (Color online) We show theoretical fits of resistance measurements in the temperature
range 5K to 50K for four films with $\delta$ = 0.0, 0.1, 0.2 and 0.4. The fits use Eq. (\ref{resistance}) in
which the carrier density used $N_2$ used in the Drude term is obtained using Eqs. (1) to (3). The inset shows the variation of $N_2$ with temperature for $\delta$ = 0.0 and $\alpha$= 0.15}
\end{figure}

We first note that this data could not be accounted for in terms of a straightforward weak-scattering
correction over a temperature-independent Drude term. This is to be expected as the
mechanism of carrier generation at the interface implies a temperature-dependent concentration.
Since several parameters are required we need to describe how have these been obtained for fitting.
The model parameters $\Delta$ and $\alpha$ are fixed from the estimates of the carrier density and
thickness of 2-DEL. They both depend, of course, on temperature and $\delta$-doping.
We have made Hall measurements (not reported here) to obtain the carrier density
in our heterostructures. Our estimates and similar estimates available in literature \cite{Thiel06}
range between 6 to 8$\times 10^{13}$/cm$^2$ for the undoped layer around 50K. The estimates for the thickness of
the electron layer are in the range 60-70 A at room temperature. We have chosen $\Delta$ and $\alpha$ so that
between 5K to 50K, $N_2$ lies between 4 to 8$\times10^{13}$/cm$^2$ and $\lambda$ ranges between 50\AA  to 10\AA. These values are shown in Table 1 for various dopings. For a typical value of parameters the inset
of Fig.\ref{Fits} shows the variation of $N_2$ with temperature.

\begin{table}
\caption{\label{Table 1} Values of model parameters used to fit resistance curves in the temperature
range 6K to 46K for $\delta$=0.0, 0.1, 0.2, 0.4 and 0.5 shown in Figure 1(d). The elastic scattering time for all
curves is $\tau_0$=2.3x10$^{-14}$s}
\begin{tabular}{ | l | l | l | l | l | l |}
  \hline
  $\delta$ & 0.0 & 0.1 & 0.2 & 0.4 & 0.5  \\
  $\alpha$ & 0.15 & 0.11 & 0.086 & 0.045 & 0.025  \\
  $\Delta (eV)$ & 0.001 & 0.001 & 0.001 & 0.001 & 0.002 \\
  $\tau _{s0} $($\times 10^{-13}$s) & 1.6 & 1.9  & 2.5 & 3.0 & 3.9 \\
  \hline
\end{tabular}
\end{table}

The value of $\tau_0$ cannot be determined unambiguously from our data, but a survey of estimates
in the literature place it to be of the order $10^{-14}$s.
We take $\tau_0=2.25\times 10^{-14}$s which then gives consistent magnitudes for other parameters.
To determine the parameters for $\tau_i$, we plot
measured $\Delta \sigma (T)= \sigma (T) - \sigma_{Drude}$ with $\ln T$. It shows an approximate linear behavior with a slope of 2.8. So we take p=2.8.
$T_0$ in the expression for $\tau_i$ is adjusted in the fitting procedure and is found to be 95K.
The first estimate for $\tau_{so}$ is taken to be $\tau_i(T_d)$ where $T_d$ is the temperature where
the increase of resistance with decreasing temperature begins to flatten. Thereafter it is fine tuned to
obtain the best fits. For the temperature range of interest we take $\epsilon_s$=300. Our fits for four resistance curves corresponding to $\delta$ = 0.0, 0.1, 0.2 and 0.4
are shown in Fig.\ref{Fits}. The fits are reasonably good in the temperatures range 6-46K with parameters shown in Table 1. A good fit is also obtained for $\delta$=0.5, but is not shown in Fig. 2.

Following comments are in order. From Table 1, we see that with doping
$\alpha$ decreases in accordance with our reasoning that Mn-doping reduces charge on LaO layer at the interface.
It is corroborated further by increase in $\tau_{so}$. Thus the layer charge density ($\alpha$) accounts for increase of resistance with doping. The gap parameter $\Delta$ does not vary with doping till $\delta$=0.4, beyond which it may rise rapidly. The quantitative accuracy of these numbers is limited due to rather approximate nature of the Thomas-Fermi theory. In the above we have
ignored the electron-electron interaction which also gives rise to $\log (T)$ terms. These are incorporated
through multiplication by a factor (1 - F) to the weak-scattering correction term \cite{Altshuler}. From
the data, we cannot disentangle this factor as here it also depends on temperature through density. This
interaction also leads to a $\log (T)$ correction to thermopower \cite{Ting}, but we find it too small to account for the data presented below. This is consistent with the remarks of Caviglia et al \cite{Caviglia10}.

From temperature of the minimum to room temperature the resistance rises rapidly by a factor of seven
or so. This model cannot account for this as this would have contributions from the usual increase of
relaxation rate as well as possible renormalization of $\phi_0$ with temperature due to charge relaxation etc.

\begin{figure}[h]
\begin{centering}
\includegraphics[width=8cm]{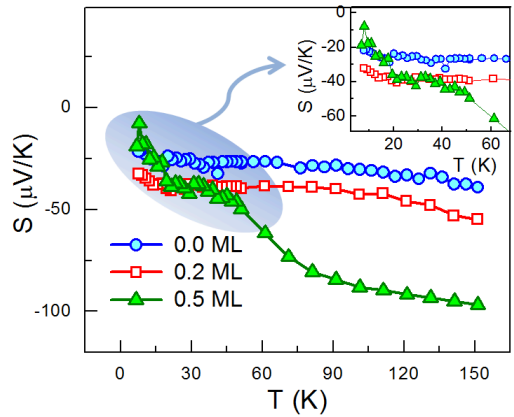}
\par\end{centering}

\caption{\label{Seebeck} (Color online) Thermopower (S) for
three  different fractions of LaMnO$_3$ layer in the temperature range 5 K to
150 K. The inset shows the differences at lower temperatures in a
zoomed plot.}
\end{figure}

Fig.\ref{Seebeck} shows the temperature dependence of thermopower
S($T$) for the LAO-STO heterostructure and two $\delta$-doped samples,
with fractions $\delta =0.2$ and $\delta=.0.5$ of monolayer of LaMnO$_{3}$ between 5 K to 150 K.
A negative TEP is observed indicating that the charge carriers are
electrons in these systems. The magnitude of S for
LAO-STO is rather small and shows a weak variation with temperature.
This is much less than what one would theoretically expect from carrier
confinement in the SrTiO$_{3}$ quantum wells. This may partly
be attributed to the effective thickness of the electron gas.
It thereby loses the two dimensional advantage. The flat temperature variation seen
here is a large departure from the free two dimensional electron gas which for example,
at a density of 2x10$^{13}$/cm$^2$ shows linear variation upto 300 K.
Note that some earlier
measurements of thermopower on LAO-STO \cite{Pallecchi,Filippetti} reported
much larger values for $ T > 77$ K. The LAO films in these studies are
much thinner (4 to 6 U.C.) and have larger sheet resistance R$_{\square}$.

The other two curves in Fig.\ref{Seebeck} show the effect of
insertion of LaMnO$_{3}$ monolayer on thermopower. At $\delta=.0.2$  one sees
a modest increase but at $\delta=.0.5$ there is a substantial increase in the
magnitude of S at temperatures higher than 60 K. At $150$
K, the magnitude of S for $\delta=.0.5$ is $96$ $\mu V/K$, which is more than
twice than that of LaAlO$_{3}$/SrTiO$_{3}$
system ($\simeq40$ $\mu V/K$). In the lower temperature range a salient feature
is the emergence of a step like variation in S. This feature could arise from
factors like: 1) due to the step like feature associated with the density
of states of the confined gas, 2) to the the phonon drag which is also expected in
this temperature range.

However, more remarkable is the behavior in the temperature range from 80 K to 150 K,
where we find that curves for the $\delta$-doped system are almost parallel.to the
undoped system. This corresponds to a temperature-independent contribution which is suggestive
of Heikes-like contribution. The mixed-valent Mn ions in Al$_{1-\delta}$Mn$_{\delta}$O$_2$ layer
are the most likely contributors to it. Beni and Chaikin \cite{Chaikin} extended
the Heikes formula to strongly-correlated systems
at high temperatures (k$_{B}$T $\gg$ $W$; $W$ is bandwidth). They showed that under
such conditions a large contribution to thermopower comes from
the spin and orbital degrees of freedom, which is basically due to entropy
associated with these degrees of freedom \cite{Chaikin,Marsh,Koshibae01}.
Thus it is natural to assume that the large enhancement in S
comes from $3d$ electrons of the Mn ions. In LaMnO$_{3}$
the Mn ions are in the $3d^{4}$ state. As argued above,
this monolayer loses some electrons in order to screen the positive field of LaO layer (see Fig. 1a).
This assumption leads to the model in which
 Mn ions in Al$_{1-\delta}$Mn$_{\delta}$O$_2$ layer exist in states of $3d^{4}$ and $3d^{3}$.
Given that the degeneracy of these states are $g_{4}$ = $10$ and
g$_{3}$ = $4$, respectively, one obtains the following formula,

\begin{equation}
S_{(Mn)}=\,\frac{k_{B}}{e}\,\ln\left(\frac{g_{4}\rho_{h}}{g_{3}\left(1-\rho_{h}\right)}\right)\label{Eq2}
\end{equation}
where $\rho_{h}$ is the number of holes from the $3d^{4}$ state.
The correct magnitude for the observed thermopower is obtained with
$\rho_{h}$ = 0.44 at $\delta$=0.5 and $\rho_{h}$ = 0.31 at $\delta$=0.2. The charge transfer per cell
implied by this is $(1-\rho_h) \delta$ which has the same trend with doping as $\alpha$. An experimental
determination of $\rho_{h}$ would be a direct way to validate the picture presented here.

To summarize, we report the transport properties of the LaAlO$_{3}$-SrTiO$_{3}$
heterostructures subjected to interface $\delta$-doping by insertion
of fractional mono-layer of LaMnO$_{3}$. This inhibits the formation of 2-DEL at the interface
reflected by increase in resistance and the larger critical thickness
of the LaAlO$_{3}$ over-layers needed to induce  2-DEL at the interface.

We have developed a simple model by extending the polarization catastrophe ideas and Thomas-Fermi
theory to make temperature-dependent estimates of the density of the electron liquid at the interface.
In this framework we provide an explanation of the above results by arguing that
Al$_{1-\delta}$Mn$_{\delta}$O$_2$ layer due
to the mixed-valent nature of Mn$^{3+}$ ion transfers charge to the LaO layer at the interface,
thereby reducing the interface field. Further by using weak-scattering mechanism we have provided
theoretical fits to the resistance data
at low temperatures and extracted the doping dependence of model parameters which substantiates the above
reasoning.
The $\delta$-doping of the interface also leads to an increase of its thermopower in the
higher temperature range, above 60 K in the present case. This increase is constant in
temperature and can be attributed to the contribution from the AlMnO$_2$ layer.
This work opens interesting possibilities for further work in this direction.
\begin{acknowledgments}
AR acknowledges the financial support from the Council of Scientific and Industrial Research (CSIR), India and Indian Institute of Technology (IIT) Kanpur and also thank V. P. S. Awana for allowing him to use PPMS for thermopower measurements. RCB acknowledges the J. C. Bose National Fellowship of the Department of Science and Technology (DST), Government of India. DK acknowledges Raja Ramanna fellowship from the Department of Atomic Energy, Government of India.
\end{acknowledgments}

\end{document}